\documentstyle[12pt,a4j]{article}
\def\lsim{\hbox{$\hskip0.3em\raisebox{-.4ex}
{$\sim$}\raisebox{.4ex}{\hskip-0.8em$<$\hskip0.1em}$}}
\def\gsim{\hbox{$\hskip0.3em\raisebox{-.4ex}
{$\sim$}\raisebox{.4ex}{\hskip-0.8em$>$\hskip0.1em}$}}
\begin{document}
\setlength{\baselineskip}{2.0pc}
\parindent = 17pt
%
%%%%%%%%%%%%%%%%%%%%%%%%%%%%%%%%%%%%%%%%%%%%%%%%%%%%%%%%%%%%%%%%%%%%%
%                             head page
%%%%%%%%%%%%%%%%%%%%%%%%%%%%%%%%%%%%%%%%%%%%%%%%%%%%%%%%%%%%%%%%%%%%%
\par
\begin{center}
\begin{large}
\begin{bf}
\vspace*{1.5cm}
Strong Coupling Phase of Chiral Gross Neveu Model
\end{bf}
\end{large}
\vskip 0.1cm
Hiroshi Nohara \par
\vskip 0.1cm
\begin{it}
Yukawa Institute for Theoretical Physics, Kyoto University, Kyoto 606,
Japan
\par
\end{it}
\vskip 0.3cm
(Received \hskip5cm )
\end{center}
\vskip 0.5cm
\par
%
%%%%%%%%%%%%%%%%%%%%%%%%%%%%%%%%%%%%%%%%%%%%%%%%%%%%%%%%%%%%%%%%%%%%%
%                              abstract
%%%%%%%%%%%%%%%%%%%%%%%%%%%%%%%%%%%%%%%%%%%%%%%%%%%%%%%%%%%%%%%%%%%%%
\centerline {\bf ABSTRACT } \par
\vskip 0.1cm
%%%%%%%%%%%%%%%%%%%%%%%%%%%%%%%%%%%%%%%%%%%%%%%%%%%%%%%%
We perform the numerical simulation of the two
dimensional chiral Gross
Neveu model using the Kogut-Susskind(KS) fermion. In the case
of SU(4), the Kosterlitz-Thouless phase transition happens
at some critical value of the coupling constant.
In the case of one flavour, there exists the strong coupling phase
in which the correlation functions vanish and the general
covariance is realized in the quantum field thoery through
 the dynamical process.
%%%%%%%%%%%%%%%%%%%%%%%%%%%%%%%%%%%%%%%%%%%%%%%%%%%%%%%%%%

\vskip 0.3cm
%\noindent
%PACS numbers: 71.27.+a, 05.30.-d, 71.10.+x, 03.65.-w
%

\newpage
%%%%%%%%%%%%%%%%%%%%%%%%%%%%%%%%%%%%%%%%%%%%%%%%%%%%%%%%%%%%%%%%%%%%%
%                              main part
%%%%%%%%%%%%%%%%%%%%%%%%%%%%%%%%%%%%%%%%%%%%%%%%%%%%%%%%%%%%%%%%%%%%%
The two dimensional Gross Neveu model had been intensively discussed
on the weak coupling phase decades ago since
it has many basic properties of renormalizable field
 theories\cite{Gr}. One of
the important problems which remain to be solved is to investigate the
phase structures in the region of the strong coupling where the
dynamically generated
mass of the fundamental fermion becomes the order of the ultra-violet
cut-off.
In the case of the usual Gross Neveu model without the classical
 chiral symmetry,
the feature of the phase for large couplings depends crucially
on the regularization. If we regularize the model by using the latticed
space time, it is shown in \cite{Co} that by adopting the KS fermion,
one can realize the phase transition of Ising type with respect to
a certain two-fermion's state and have the strong
coupling phase where there is the chiral symmetry.

On the other hand, there are evidences for a strong coupling phase
of the chiral Gross Neveu model from several points of view.
According to the analysis based on the Bethe Ansatz \cite{An} the effective
coupling constant which is considered to be the
 coefficient of the interaction
between the composite fermions becomes infinite at a certain
 finite value of the
coupling constant of the Gross Neveu model. Above this value, the
effective coupling constant becomes negative which implies
 that the interaction is the attractive force and that the system is
not in a physical state because of the instability. Somehow
it is impossible to apply the Bethe Ansatz approach to the model
in a whole region as it is.
Also the results of the  perturbative
calculations of the beta function suggest the existence of the fixed point
since the position of it does not change so drastically by including
the correction terms order by order \cite{So}.
{}From these facts, we are inclined to think
that there is much possibility that the model is in
some new phase for large couplings.
It is well known that in the spectrum of the Hamiltonian for weak couplings,
there are massive ones and a massless one which has its origin in the chiral
symmetry. An interesting question is to ask
the spectrum of the Hamiltonian for large values of the coupling, for example,
if a massless particle exists or not since the massive states
mentioned before will be
decoupled from the system as indicated by the large fermion mass.

In this article, we discuss the results of the numerical simulation of
the chiral Gross Neveu model using the KS fermion defined
 on the lattice the size
of which is 16 by 16. The numerical data tell us that in the case of
SU(4), there is a certain phase transition and exists the strong
coupling phase in which the order parameter becomes massive, which can
be also derived from the argument based on the 1/N expansion analysis. On
the other hand, in the case of one flavour, there is no such phase transition
as the previous example and the model shows the triviality indicated by
the vanishing of one and two point functions of the order parameters in
the region of large couplings. We will try to describe this state in terms
of the unbroken phase of the general covariance.

The lattice model to describe the chiral SU(N) Gross Neveu model would
be given by the following Lagrangian,
\begin{equation}
L=\sum_{i=1}^{\mbox{N}}\overline{\psi_{i}}
\left( \triangle_{\mu} \gamma_{\mu}\otimes \mbox{1}
-\frac{1}{2}\gamma_{3}\otimes \gamma_{\mu}\gamma_{3}\Box_{\mu}
+g\left(1\otimes \sigma +i \gamma_{3}\otimes \pi \right)
\right) \psi_{i}+\frac{\mbox{Tr}(\sigma^{2}+\pi^{2})}{2} , \label{kslag}
\end{equation}
where $\triangle_{\mu}$ and $ \Box_{\mu}$ are the lattice differentials of
the first and second order respectively and $\gamma_{\mu}(\mu=1,2),\gamma_{3}$
are gamma matrices in two dimensions,
 $\{\gamma_{\mu},\gamma_{\nu}\}=2\delta_{\mu \nu}
,i \gamma_{3}= \gamma_{1}\gamma_{2}$. $\psi_{i}$ are the lattice
fermions and have two internal indices,
those of the spin and the flavour besides the suffix $i$.
Residual fields ${\bf \sigma}$ and ${\bf \pi}$ have two diagonal components
$\sigma_{k}$ and $\pi_{k}
(k=1,2)$ respectively which operate on the space of the coordinate
 of the space time and
 the flavour. They are expressed as
$\sigma =\sigma_{1}(1+\gamma_{3})/2+\sigma_{2}(1-\gamma_{3})/2$,
$\pi =\pi_{1}(1+\gamma_{3})/2+\pi_{2}(1-\gamma_{3})/2$.

This model is invariant under the rotation with respect to $\sigma$ and $\pi$,
\begin{equation}
1 \otimes \sigma +i \gamma_{3}\otimes \pi \rightarrow e^{i \theta \gamma_{3}
\otimes \gamma_{3}}(1\otimes \sigma +i \gamma_{3}\otimes \pi ) \label{sym2}.
\end{equation}

Note that there is the quasi-chiral symmetry at $g=0$,
\begin{equation}
\psi_{i} \rightarrow e^{i \theta\gamma_{3}\otimes  \gamma_{3}}\psi_{i}
 \label{sym1}.
\end{equation}

Unfortunately, it is very difficult to treat (\ref{kslag}) from a
technical point of view. Actually for generic values of ${\bf \sigma}$ and
${\bf \pi}$, the determinant of the operator acting on fermions is
complex valued
and there is no practical method to calculate its phase in the case that
the size of matrix is over 1000 by 1000. Instead of (\ref{kslag}), we consider
the one
\begin{eqnarray}
L&=&\sum_{i=1}^{\mbox{N}/2} \overline{\psi_{i}}
\left( \triangle_{\mu} \gamma_{\mu}\otimes \mbox{1}
-\frac{1}{2}\gamma_{3}\otimes \gamma_{\mu}\gamma_{3}\Box_{\mu}
+g\left(\sigma +i \gamma_{3}\pi \right)
\right) \psi_{i}+\sigma^{2}/2+\pi^{2}/2 , \\ \nonumber
&=&\sum_{i=1}^{\mbox{N}/2}\overline{\psi}_{i}{\cal D}\psi_{i}
+\sigma^{2}/2+\pi^{2}/2, \label{kslag1}
\end{eqnarray}

which is obtained by simply replacing $\sigma$ and $\pi$ with
the scalar function $\sigma$ and $\pi$ operating as identities in the space
of the flavour. In this way, we have the real valued determinant as there
is the representation in which $\gamma_{\mu}(\mu =1,2)$ and $i \gamma_{3}$
are all real.
The drawback of the second model is that the global symmetry (\ref{sym2})
is broken by the terms of the second derivative which are of the first order
of the lattice constant. We assume that if the size of the lattice is large
enough, these terms are negligible
and that these two models are same in essence.

The algorithm we adopt for the simulation is the Hybrid algorithm
which is suitable for treating a dynamical fermion in general \cite{Du}.
 Let us give a short explanation
of the method. In our model, it is necessary to generate the configuration of
$\sigma$ and $\pi$ with the weight proportional to $e^{-S}$ where $S$ is
the effective action for $\sigma$ and $\pi$,
\begin{equation}
S=\sum_{n} ( \sigma^{2}/2+\pi^{2}/2)(n)
-\frac{\mbox{N}_{f}}{4}\mbox{Trln}{\cal D}^{\dagger}{\cal D},\label{wei}
\end{equation}
$n$ is the label of the site on the lattice and
the last term comes from the integration of the fermions.
To this end, we take the following steps. 1) Generating the set $\{ \sigma(n)
,\pi(n) \}$ in some appropriate way.
2) Choose the set$\{p_{\sigma}(n),p_{\pi }(n)\}$
 from the ensemble of the Gaussian random number $\{ \xi_{n} \}$ satisfying
\begin{equation}
<\xi_{n}>=0,<\xi_{n}\xi_{m}>=\delta_{nm}. \label{rand}
\end{equation}
3)Iterating the differential equation below for suitable time steps
 $N_{md}\delta t$
\begin{eqnarray}
&&{\cal H}=\sum_{n} (p_{\sigma}^{2}/2+p_{\pi}^{2}/2)(n)+S,\nonumber\\
&&\dot{\sigma}(n) =p_{\sigma}(n),\dot{\pi}(n)=p_{\pi}(n),
\dot p_{\sigma}(n)=-\partial {\cal H}/\partial \sigma(n),
\dot p_{\pi}(n)=-\partial {\cal H}/\partial \pi (n), \label{eqmo}
\end{eqnarray}
where the new variables $p_{\sigma}$ and $p_{\pi}$ are considered to be
 conjugate to $\sigma$ and $\pi$ respectively.
4) Restore the configuration of ${\sigma}$ and ${ \pi}$ at the final
step.
5) Go back to 2), and repeat 2) to 4) until we have enough data.

We remark that in order to apply this algorithm to our model, it is necessary
that $S$ is real and that $\mbox{Det} {\cal D}$ has a definite sign.
The first one is clearly satisfied if the second one is.
It can be checked easily that the determinant is positive for
all configurations of $\sigma$ and $\pi$ in the case of any size
 of the lattice.
To be rigid, we can treat the case of the even number of
$\mbox{N}_{f}$
 but assume that
the weight (\ref{wei}) is correct for odd cases.

In 3) we perform the iteration using the leap-frog method \cite{Go}
 and estimate
that the error is of order $(\delta t)^{2} N_{md}$.
The number of the iteration $N_{md}$ in 3) is determined so that
 each data acquires most independence.
Actually we calculate autocorrelation function $f(t)=<{\cal O}(t)
{\cal O}(0)>$ and its decay constant $T$ defined by $f(t)\sim e^{-t/T}$
, that is, the autocorrelation time.
The optimum value of $\mbox{N}_{md}$ would give the minimum value of
$T$ in general. In our simulation, $T$ simply decreases as far as
we observe ($60 \lsim N_{md} \lsim 120$) and $N_{md}$
is chosen to be 100 and the unit of the
time step $\delta t$ to be 0.01 to save
the time of CPU. We checked that the result of the simulation does
not depend on $N_{md}$ around this value and will have the
reasonable data as one sees later. We also comment that
the number of the stored data is about several thousand.

The most difficult thing in the above process is to calculate the
trace of the operators
$
{\cal D}
({\cal D}^{\dagger}{\cal D})^{-1}
\partial {\cal D}^{\dagger}{\cal D}
/\partial \sigma (n)
({\cal D}^{\dagger}{\cal D})^{-1}{\cal D}^{\dagger}
$
and
$
{\cal D}
({\cal D}^{\dagger}{\cal D})^{-1}
\partial {\cal D}^{\dagger}{\cal D}
/\partial \pi (n)
({\cal D}^{\dagger}{\cal D})^{-1}{\cal D}^{\dagger},
$
which appear in the equation of motion (\ref{eqmo}) since it is
necessary
 to make
the inverse of ${\cal D}^{\dagger}{\cal D}$ many times and it costs a lot
of time. Instead of doing it, we adopt the noisy estimator as follows
 \cite{Du}.
In each step in 3), we estimate the matrix element $\xi {\cal O} \xi$ where
$\xi$ is the vector each component of which is the newly refreshed
normalized Gaussian random number(\ref{rand}). If we set
$\delta t$ and $N_{md}$ to be small and large enough respectively, this
procedure would give the nice solution to the equation.

Now we mention the result of the simulation. The simulation is
performed under the boundary condition that the fermions are periodic
and anti-periodic in the direction of the space ($x$) and time ($t$)
respectively in order to remove the zero mode of the lattice differentials.
Firstly we discuss the case of SU(4).
An interesting quantity is the correlation function of the order parameter
$<\sigma_{+}(t) \sigma_{-}(0)>$ ($\sigma_{+}=\sigma +i \pi$,
$\sigma_{-}=\sigma -i \pi$).
 In Fig. \ref{su4cor}, we will show the numerical data for $g=0.5,0.8,0.9$.
The fitting curves indicated by solid ones are determined by the least square
method.
Taking account of these data we conclude
that there is a certain critical value of coupling constant
$g_{c}(\sim 0.8)$ and for
$g < g_{c}$ the correlation function behaves like $\sim t^{\alpha}$ and
for $g > g_{c}$, it decays exponentially $\sim e^{-mt}$ where the mass
$m$ becomes zero at $g_{c}$. Qualitatively
 $\alpha$ decreases and $m$ increases as the function of $g$.

The behaviour of the correlation function mentioned above is reminiscent
of the two dimensional XY model. Indeed, according to the argument
 similar to the one \cite{Af},
the leading term in the 1/N expansion of the effective action $S$ consists
of the Hamiltonian of the XY model and higher derivative terms. Provided
that there exists the phase transition of Kosterlitz-Thouless type,
the effective action reasonably approximates the Hamiltonian of the XY
 model around the
critical point since the higher derivative terms can be neglected.
We should comment that if we use another lattice fermion, for example,
the Wilson fermion, things change for strong couplings.
\footnote{We
thank Y. Kikukawa for suggesting this model.}
Let us consider  the model, for example,
\begin{equation}
L=\sum\overline{\psi_{i}}
\left( \triangle_{\mu} \gamma_{\mu}\otimes \mbox{1}
+\frac{1}{2}1 \otimes \gamma_{1}\Box_{\mu}
+g (\sigma +i \gamma_{3} \pi )\otimes (1+\gamma_{3})/2
\right) \psi_{i}+{\bf \sigma}^{2}/2+{\bf \pi}^{2}/2, \label{wil}
\end{equation}
where $(1+\gamma_{3})/2$ projects only the first flavour.
We assume that (\ref{wil}) is equivalent to the one which consists of
the free fermion and the interacting one.
This model
has two global symmetries (\ref{sym1}) (\ref{sym2}) mentioned before
and has the real valued determinant. We checked that the correlation
function shows a power like
behaviour up to $g=1.1$ and conjecture that there is no phase transition.

Now we turn to the case of only one flavour.
The chiral Thirring model can be solved easily by using bosonization
 in the region of small couplings and has only massless bosonic particle
 in the spectrum. To be concrete, let us consider the following Lagrangian
\begin{equation}
L=\overline{\psi}\gamma_{\mu} \partial_{\mu} \psi -
 \lambda^{2}\psi^{\dagger}_{R} \psi_{L}\psi^{\dagger}_{L}\psi_{R},
 \label{Thir}
\end{equation}
where $\lambda$ is related to $g$ in (\ref{kslag1}) as $\lambda^{2} =2
g^{2}$. The bosonization
rule is that $\psi$ is described by the free scalar field $\phi=
\phi_{L}+\phi_{R}$ as
$\psi_{L,R}=e^{i \phi_{L,R}}$ where $\partial_{L,R}\phi_{R,L}=0$.
According to it, the Lagrangian  is rewritten in terms of $\phi$ as
\begin{equation}
L=(1-\lambda^{2})\partial_{\mu}\phi\partial_{\mu}\phi.\label{boson}
\end{equation}
{}From this fact the correlation functions are
\begin{equation}
<\psi^{\dagger}_{R}(t)\psi_{R}(0)>\propto t^{-1/(1-\lambda^{2})},\ \ \
<\sigma_{+}(t)\sigma_{-}(0)> \propto t^{-2/(1-\lambda^{2})}.\label{cf1}
\end{equation}
Note that the above argument is valid for the case $|\lambda |< 1$ as
seen from the fact that for
$|\lambda |> 1$, the Lagrangian has the wrong sign because of the
 coefficient
$1-\lambda^{2}$ while the Lagrangian for the fermion is well
 defined for arbitrary
values of $\lambda$. Above the critical value of $\lambda_{c}=1$,
no analytic solution is known so far.

This case is more difficult
to treat in the simulation compared with the previous one since
the speed of the convergence is very slow.
In general the estimation of the correlation function of the
order parameter is almost impossible for the values $g \lsim 0.8$.
To overcome this problem
we use the model which is obtained by eliminating $\pi$ from (\ref{kslag1}).
(The coupling constant in the Lagrangian on the lattice is identified
with $\lambda$ in (\ref{Thir}) since in the lattice regularization,
the coupling term $g\overline{\psi}\psi(n)\sigma(n)$ can be identified with
$-g^{2}\psi^{\dagger}_{R} \psi_{L}\psi^{\dagger}_{L}\psi_{R}(n)$
using the equation of motion.)
This model enables us to calculate the correlation functions
of the fermion which are easier to estimate since the power of them is
half of that of the order parameter as we see from (\ref{cf1}).
In the model we treated so far, it is very difficult
to obtain nice results of the correlation function
because it has ``approximately'' the continuous symmetry (\ref{sym2})
 and it would take much time to sum over the phase.

 The particular feature of
the data is that for the couplings $\lambda \gsim 1$, the correlation
functions of the fermion and order parameter vanish on the contrary
to the previous case. On the other hand, for $\lambda \lsim 1$
 that of the
fermion behaves as $\sim t^{\beta}$. In Fig. \ref{u1cor}, we see
 that the power
$\beta$ for $\lambda=0.5$ is a little smaller than the one obtained
from the above analysis and
that the correlation function of the order parameter
 for $\lambda=1.2$
 is of the order of the square of its expectation value
which should vanish because of the chiral symmetry.

Let us concentrate on the case of $\lambda =\lambda_{c}$.
If we define the Thirring model to be (\ref{boson}) as the limit of
$\lambda \rightarrow \lambda_{c}-0$, it follows that the model
 is trivial
in the sense that all the correlation functions of the fermion
 vanish since
the power in (\ref{cf1}) is infinite and there is no dynamical degree
of freedom. Moreover as the Lagrangian (\ref{boson}) is zero, the
partition function is
\[ Z=\int_{M} {\cal D}\phi e^{-S} = (2\pi)^{\mbox{Area}(M)},\]
and depends only on the area of the space time.
%This is interpreted in
%terms of the fermion as follows. The Lagrangian (\ref{Thir}) is
% invariant under the
%local $GL(1,C)$ transformation, $\psi \rightarrow G\psi$,
%$\overline{\psi} \rightarrow \overline{\psi}G^{-1}$ where $g$ is
 the element
%of the group. Indeed if we regularize the each term of the Lagrangian by
%the point splitting, only the irrelevant terms which include polynomials
%$G$ and its derivatives are added to it after the transformation.
%Thanks to this symmetry, we can gauge the fermion so that it is simply
%a constant real field which makes the Lagrangian zero as easily seen.
If we define the model
on the space time $M$ with the metric $g_{\mu \nu}$ in general,
 the expression
of $Z$ is the same as the above one. From this fact, at
$\lambda=\lambda_{c}$
we have the phase which does not depend on the back ground metric at all.

As long as we compare the correlation function at $\lambda_{c}$ with those
for $\lambda > \lambda_{c}$, there is no difference detected through the
numerical method we performed. We conclude that in the Thirring
model the correlation function does not depend on the distance between
the inserted operators above $\lambda_{c}$ and the model lives in the phase
where the general covariance is not broken.

It would be very interesting
that this strong coupling phase also has the symmetry we saw at
 $\lambda_{c}$ and is
the trivial phase where there is no propagating particle and the system
is not sensitive to the geometrical data of the space time.
The point is that this type of phase transition could happen
through the dynamical process. We would like to expect that this type
of phenomena
is realized in a more realistic model, for example, the unification theory
which includes the four dimensional quantum gravity.

The calculation is performed by using the work station system at Yukawa
Institute and facom.m1800 at Kyoto-University.
We thank Y. Kikukawa for useful discussions and K. Ikeda, A.
Nakamichi, M. Miyamoto, S. Yosida and S. Takasugi
for continuous encouragement.

\newpage
%%%%%

%%%%%%%%%%%%%%%%%%%%%%%%%%%%%%%%%%%%%%%%%%%%%%%%%

\newpage
%%%%%%%%%%%%%%%%%%%%%%%%%%%%%%%%%%%%%%%%
%**********************************************************
%************** F I G U R E   C A P T I O N S *************
%**********************************************************

%\figure{ \label{aucor}}
%The autocorrelation time's dependence on the number of the
%iteration $N_{md}$ for $g=0.5$ and $\mbox{N}_{f}$=4.

\begin{figure}
\caption{The correlation functions $<\sigma_{+}(t)\sigma_{-}(0)>$ for SU(4).
 The values of the coupling constant and the fitting curves
 will be mentioned in order. a) $g=0.5$,
$y=0.28 t^{-0.68}$. b) $g=0.8$, $y=0.17(t-8)^{2}+20.6$.
c) $g=0.9$, $y=19 e^{-0.085(t-1)}$. We take an average over the space
variable in b) and c).  In b) the fact that the fitting curve is the parabola
means the existence of the free massless bosonic particle.}
\label{su4cor}
\end{figure}

\begin{figure}
\caption{The correlation functions for the Thirring model.
These are calculated in the model without $\pi$.
 a) $<\psi_{R}^{\dagger}(t)\psi_{R}(0)>$ for $\lambda=0.5$. The
power of the fitting curve is 1.1 and the
theoretical value is $4/3 \sim 1.3$.
b) The correlation function $<\sigma(t)\sigma(0)>$ for $\lambda=1.2$.
The solid line is the upper bound of $<\sigma >^{2}$.}
\label{u1cor}
\end{figure}
%%%%%%%%%%%%%%%%%%%%%%%%%%%%%%%%%%%%%%%%%%%%%%%%%%%%%%%%%%%%%%%%%%%%%
\end{document}